\documentclass[graybox]{svmult}
\usepackage{mathptmx}       % selects Times Roman as basic font
\usepackage{helvet}         % selects Helvetica as sans-serif font
\usepackage{courier}        % selects Courier as typewriter font
\usepackage{type1cm}        % activate if the above 3 fonts are
\usepackage{makeidx}         % allows index generation
\usepackage{graphicx}        % standard LaTeX graphics tool
\usepackage{multicol}        % used for the two-column index
\usepackage{siunitx}
\usepackage{lineno}
\makeindex             % used for the subject index
\begin{document}
\title*{J/$\psi$ production as a function of charged-particle multiplicity in pp collisions at $\sqrt{s}$ = 5.02 TeV with ALICE}
\titlerunning{Anisa Khatun}
\author{\large Anisa Khatun (for the ALICE collaboration)\\
               Department of Physics, Aligarh Muslim University, Aligarh\\
        \email {anisa.khatun@cern.ch}}
\authorrunning{J/$\psi$ Vs Multiplicity at pp 5.02 TeV} 
\maketitle
\abstract{The relative J/$\psi$ yields as a function of relative charged-particle multiplicity in pp collisions at $\sqrt{s}$ = 5.02 TeV, measured at forward rapidity, are investigated for the first time. A linear increase of the relative J/$\psi$ yield with respect to multiplicity is observed. A comparison of the findings of the present work with the available ALICE measurements obtained in pp collisions at $\sqrt{s}$ = 13 TeV at forward and mid-rapidity indicates that the increase of J/$\psi$ with multiplicity is independent of energy but exhibits a strong dependence on the rapidity gap between the J/$\psi$ and multiplicity measurements. The results are compared with theoretical model calculations.
}
\vspace{15pt}
%\begin{keyword}
%\textbf{Keywords:} J/$\psi$, multiplicity, pp collisions
%\end{keyword}
%
%
%
%
%\newpage
\section{Introduction}
The event by event multiplicity of charged-particles ($N_{\rm ch}$) produced in high energy collisions is taken as a simple yet important variable for understanding the collision dynamics. The multiplicity measurement is useful for studying the general properties of particle production. The quarkonium production as a function of charged-particle multiplicity (${\rm d}N_{\rm{ch}}/{\rm d}\eta$) in proton-proton (pp) and proton-nucleus (p-Pb) collisions is considered as an interesting observable to understand multi-parton interactions (MPI) and to explore the presence of collective behavior in small systems. Such studies can play an important role in understanding the production mechanism of heavy quarks from hard processes, and its relation with soft scale processes~\cite{AliceA}. The multiplicity dependence of J/$\psi$ production has been investigated in pp collisions at $\sqrt{s}$ = 7 and 13 TeV and p-Pb collisions at $\sqrt{s_{\rm NN}}$ = 5.02 TeV at forward and mid-rapidity using the ALICE detector~\cite{Alice2,Alice13tev,JMB}. A similar study has also been carried out for D mesons~\cite{Alice3}. In all cases, an increase of the relative particle yields as a function of the relative charged-particle multiplicity (${\rm d} N_{\rm ch}/{\rm d} \eta/<{\rm d} N_{\rm ch}/{\rm d} \eta>$) has been reported. In these proceedings, we report on the study of the multiplicity dependent relative J/$\psi$ yield (${\rm d}N_{J/\psi}/{\rm d}\eta/<{\rm d}N_{J/\psi}/{\rm d}\eta>$) at forward rapidity in pp collisions at $\sqrt{s}$ = 5.02 TeV, which has not been measured until now. The findings are compared to the results obtained in pp collisions at $\sqrt{s}$ = 13 TeV to explore the energy and rapidity dependences of the correlation between soft and hard physics processes. The results are also compared with theoretical model calculations. 
\section{Experimental setup and analysis strategy}
In ALICE~\cite{AliceC}, the charmonia are measured via their dilepton decay channels. The Muon Spectrometer is used for the reconstruction of muons coming from J/$\psi$ decays at forward pseudorapidity (-4 $< \eta <$ -2.5) while the central barrel detector covers the mid-pseudorapidity range ($|\eta|<$ 0.9) and measures J/$\psi$ in the di-electron decay channel. The central barrel detector system includes the Inner Tracking System (ITS) and the Time Projection Chamber. The ITS consists of six layers of silicon detectors. The Muon Spectrometer consists of Muon Tracking Chambers, used for tracking the muons and the Muon Trigger system that allows the selection of collisions that contain opposite sign dimuon pairs. %T0 and V0 satisfy triggering condition in the forward rapidity region. V0 detector is also used for multiplicity estimation covers pseudorapidity ranges  -3.7 $< \eta <$ -1.7  and   2.8 $< \eta <$ 5.1.%
Multiplicity is described in terms of tracklets, which are reconstructed by pairs of hits in the two innermost ITS layers, the Silicon Pixel Detector (SPD). The analysis is restricted to the event class $\rm INEL>0$ defined by requiring at least one charged-particle produced in $|\eta|<$1. %In total 1.2 million dimuon events and 10.6 million Minimum Bias(MB) events are analyzed.% 
Several selection criteria are applied to make sure that both the SPD vertex position and the event charge-particle multiplicity are properly determined. Pile-up events are removed~\cite{Alice2}. The relative charged-particle multiplicity in the $i\textsuperscript{th}$ multiplicity range, estimated at mid-rapidity ($|\eta|<$1) and for $\rm INEL>0$ events, is calculated using the following formula:
%The \avgdnchdeta is calculated in the $i^{th}$ multiplicity bin at mid-rapidity range for $INEL>0$ event class, using the following definition:
%
\begin{equation}
\label{eqn:dnchdeta}
%\centering
\frac{{\langle \rm d}N_{\rm{ch}}/{\rm d}\eta \rangle^{ i}}{\langle {\rm d}N_{\rm{ch}}/{\rm d}\eta \rangle} =  \frac{ f(\langle N_{\rm trk}^{\rm corr} \rangle^{ i})}{\Delta \eta \cdot \langle {\rm d}N_{\rm{ch}}/{\rm d}\eta \rangle_{\rm INEL>0} }
\end{equation}
Where $\langle N_{\rm trk}^{\rm corr} \rangle^{i}$ is the average number of SPD tracklets, corrected by acceptance and efficiency in the multiplicity bin $i$. The correlation function $f$ is evaluated from Monte Carlo simulations to convert the corrected number of tracklets into a number of primary charged-particles produced within $|\eta|<$1. The relative J/$\psi$ yield is estimated in the $i\textsuperscript{th}$ multiplicity bin by using the following equation:
\begin{equation}
\label{eqn:jpsiyield}
%\centering
\frac{{ \rm d}N^{ i}_{{\rm J}/\psi}/{\rm d}y }{\langle {\rm d}N_{{\rm J}/\psi}/{\rm d}y \rangle} =  \frac{N^{i}_{{\rm J}/\psi} }{N_{{\rm J}/\psi}}\times\frac{N_{\rm{MB}}}{N^{ i}_{\rm{MB}}}\times \epsilon
\end{equation}
Where $N_{{\rm J}/\psi}$ and $N_{\rm{MB}}$ are the number of J/$\psi$ and MB events respectively. The factor $\epsilon$ is a combination of several corrections that account for trigger selection, event selection, SPD vertex quality assurance and pile-up rejection. Details of the analysis procedure can be found in Ref.~\cite{JMB}.
%
%\begin{figure}[h!]
%\centering
%\subfigure{\includegraphics[width=0.46\linewidth]{2018-May-09-JpsiSignal_pp5.eps}}
%\subfigure{\includegraphics[width=0.45\linewidth]{2018-May-08-Ncharge-CorrectedNtraklet.eps}}
%\caption{ \footnotesize{(Right)Relative J/$\psi$ yield as a function of the relative charged-particle density measured at forward rapidity in pp collisions at $\sqrt{s}$ = 5.02 TeV.(Left) Comparison of results at $\sqrt{s}$ = 5.02 TeV with forward rapidity and mid-rapidity 13 TeV results. The dotted line represents to a linear correlation. Bottom panel: ratio of Relative J/$\psi$ yield to relative charged-particle density as a function of multiplicity.}}
%\label{fig:peformplot}
%\end{figure}
%
\section{Results and discussions}
\begin{figure}[h!!]
\centering
\includegraphics[width=0.8\linewidth]{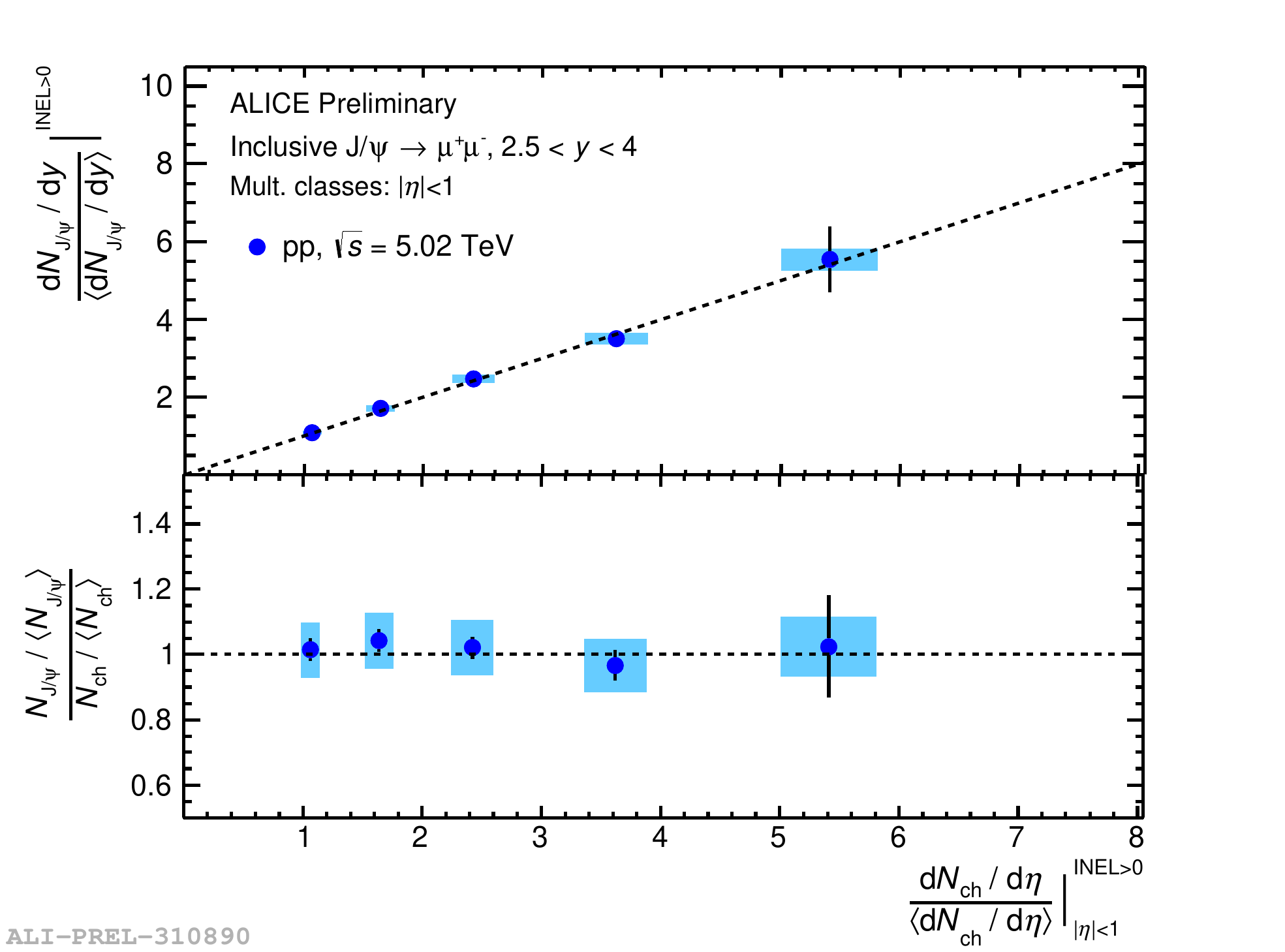}
%\subfigure{\includegraphics[width=0.46\linewidth]{2018-09-30-JpisVsMult_ForwardMidRapidity_comparison.pdf}}
\caption{\footnotesize{ The relative J/$\psi$ yield as a function of the relative charged-particle density measured at forward rapidity in pp collisions at $\sqrt{s}$ = 5.02 TeV. Bottom panel: ratio of the relative J/$\psi$ yield to the relative charged-particle density as a function of multiplicity.}}
\label{fig:Jpsimulpp5}
\end{figure}

\begin{figure}[h!]
\centering
\includegraphics[width=0.8\linewidth]{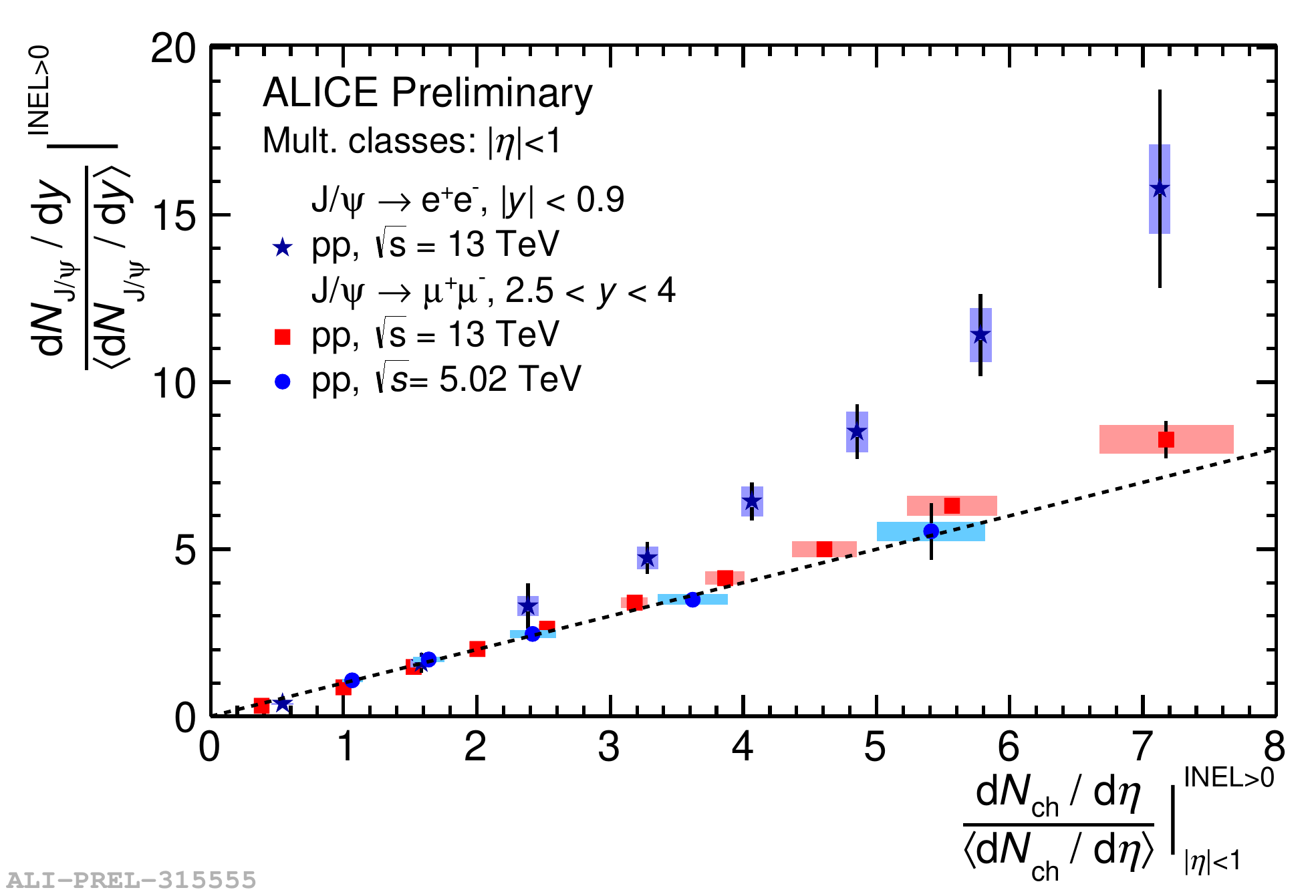}
\caption{Comparison of the relative J/$\psi$ yields as a function of the relative charged-particle multiplicity at forward rapidity in pp collisions at $\sqrt{s}$ = 5.02 TeV and 13 TeV and at mid-rapidity at $\sqrt{s}$ = 13 TeV.}
\label{fig:jpsirap}
\end{figure}

\begin{figure}[h!]
\centering
\includegraphics[width=0.8\linewidth]{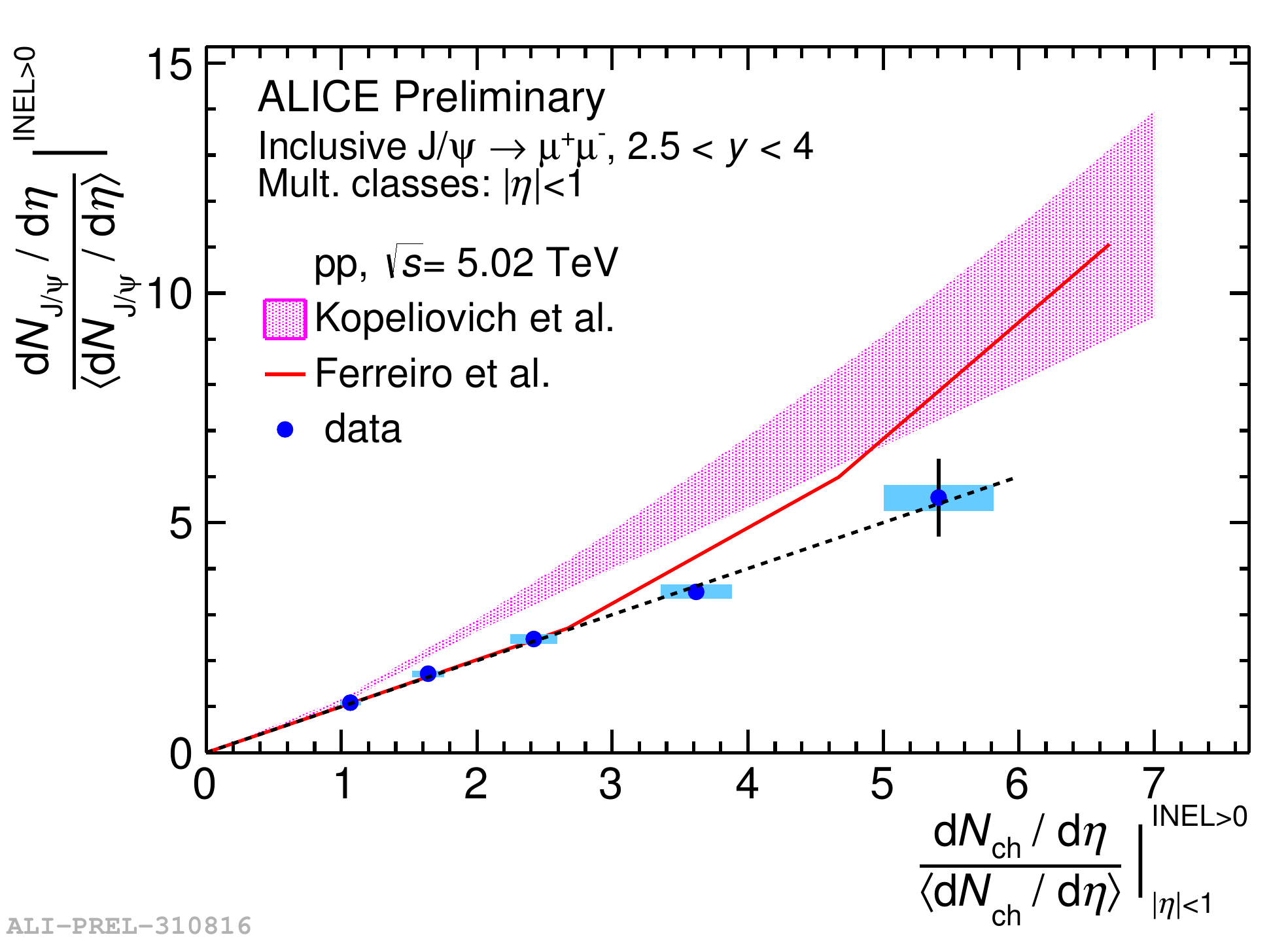}
\caption{\footnotesize{Theoretical models compared to the relative J/$\psi$ yield at $\sqrt{s}$ = 5.02 TeV.}}
\label{fig:jpsimodel}
\end{figure}

The multiplicity dependence of the J/$\psi$ yield is displayed for pp collisions at $\sqrt{s}$ = 5.02 TeV in Fig.~\ref{fig:Jpsimulpp5}. The correlation between multiplicity and J/$\psi$ is compared to a linear function with a slope of 1 (y=x). The ratio of the relative J/$\psi$ yield to this diagonal as a function of multiplicity is displayed in the bottom panel of Fig.~\ref{fig:Jpsimulpp5}. %The ratio shows no deviation from unity at forward rapidity. 
It is interesting to note that the ratio is consistent with unity over the full multiplicity range. This, in turn implies that the production of J/$\psi$ scales linearly with the underlying event activity when a gap is present between the rapidities at which ${\rm d}N_{\rm{ch}}/{\rm d}\eta$ and J/$\psi$ are measured.
The J/$\psi$ yield observed at $\sqrt{s}$ = 5.02 TeV is compared with those reported for forward and mid-rapidity ALICE measurements in pp collisions at $\sqrt{s}$ = 13 TeV and the results are displayed in Fig.~\ref{fig:jpsirap}. %No strong change of multiplicity dependence with center of mass energy is observed so far at forward rapidity. At mid-rapidity multiplicity study by ALICE, both multiplicity and J/$\psi$ are measured in mid rapidity i.e. no y-gap is provided. The comparison with mid-rapidity shows stronger than linear increment towards higher multiplicity. The rapidity dependence may cause due to the reason for autocorrelations or jet biases~\cite{auto}.\\%
%It is evident from the figure that J/$\psi$ yield estimated for different multiplicity bins are almost independent of the beam energy in the forward rapidity region.%
It is observed that the relative J/$\psi$ yield estimated for different multiplicity bins are practically independent of the beam energy in the forward rapidity region. However in the mid-rapidity region for $\sqrt{s}$ = 13 TeV data, the measured relative J/$\psi$ yield increases faster than linearly with increasing multiplicity. Such a dependence may be attributed to the presence of auto-correlations between the J/$\psi$ and the multiplicity measurements at mid-rapidity, for instance because of a possible jet bias~\cite{auto}.
%
%Such a dependence of \jpsi yield on relative multiplicity may be envisaged due to the reason for autocorrelations or jet biases~\cite{auto}. 
%
%
%
%
%The experimental results which are obtained from ALICE data at $\sqrt{s}$ = 5.02 TeV, are compared with theoretical model calculations is plotted in \textbf{Figure~\ref{fig:jpsimodel}}. Theoretical model are provided to explain the experimental data from Kopeliovich et al.~\cite{Alicemodels4} and from Ferreiro et al. (the percolation model) ~\cite{Alicemodels1} at forward rapidity. 
%In order to compare the findings of the present work with the predictions of theoretical models, values of J/$\psi$ yield against multiplicity are plotted in \textbf{Fig.~\ref{fig:jpsimodel}}. Variations of J/$\psi$ yield with multiplicity at forward rapidity predicted by models proposed%
%
In Fig.~\ref{fig:jpsimodel} the relative J/$\psi$ yields as a function of multiplicity are compared to predictions from two theoretical models proposed by Kopeliovich et al.~\cite{Alicemodels4} and Ferreiro et al. (the percolation model) ~\cite{Alicemodels1}. The percolation string model shows a linear increase at low density and quadratic increase at higher density and describes the data well at low multiplicity. The Kopeliovich model takes into consideration the contributions of higher Fock states to reach high multiplicities in pp collisions. As a result of a higher number of gluons, the J/$\psi$ production rate is also enhanced. A stronger than linear increase with multiplicity is observed. 
\section{Conclusions and outlook}
The multiplicity dependence of J/$\psi$ has been studied in pp collisions at $\sqrt{s}$ = 5.02 TeV. A linear increase of the relative J/$\psi$ yield is observed as a function of multiplicity at forward-rapidity. The multiplicity dependence is independent of $\sqrt{s}$ at LHC energies. The increase of J/$\psi$ production seems to depend on the rapidity gap between the J/$\psi$ and the multiplicity measurement. The finding reveals too that the percolation model describes the data well at forward-rapidity for relative multiplicities $<$ 3. The multiplicity dependent study of various quarkonium states will shed more light on this topic.
\section*{Acknowledgments}
We are thankful to UGC and DST for financial support.

\end{document}